\documentclass[aps,pra,showpacs,twocolumn,groupedaddress,floatfix,longbibliography]{revtex4-1}
\usepackage[utf8]{inputenc}
\usepackage[T1]{fontenc}
\usepackage{amsfonts}
\usepackage{amssymb}
\usepackage{amsmath}
\usepackage{graphicx}
\usepackage{epsfig}
\usepackage{color}
\usepackage{xcolor}
\usepackage{tikz}
\usepackage{amsmath,bm}
\usepackage{booktabs}
\usepackage{placeins}
\usepackage[colorlinks=true, linkcolor=blue, citecolor=blue, urlcolor=blue]{hyperref}

\DeclareUnicodeCharacter{754F}{\ensuremath{\eta}}
\DeclareUnicodeCharacter{5473}{\ensuremath{\zeta}}
\DeclareUnicodeCharacter{03B7}{\ensuremath{\eta}}
\DeclareUnicodeCharacter{03B6}{\ensuremath{\zeta}}
\DeclareUnicodeCharacter{03BB}{\ensuremath{\lambda}}
\DeclareUnicodeCharacter{9225}{--}
\DeclareUnicodeCharacter{64DF}{C}
\definecolor{orcidlogocol}{HTML}{A6CE39}
\DeclareRobustCommand{\orcidicon}[1]{\href{https://orcid.org/#1}{\tikz[baseline=-0.6ex]\node[fill=orcidlogocol,circle,inner sep=0.25pt,minimum size=1.45ex]{\scriptsize\textcolor{white}{\sffamily iD}};}}

\input{tcilatex}
\begin{document}

\title{Exact interlayer triplet-pairing eigenstates in the extended Hubbard
model}
\author{F. X. Liu\,\orcidicon{0009-0008-9468-8972}}
\author{Z. Song\,\orcidicon{0000-0002-3315-4589}}
\email{songtc@nankai.edu.cn}
\affiliation{School of Physics, Nankai University, Tianjin 300071, China}

\begin{abstract}
$\eta$-pairing symmetry generalizes the pairing mechanisms in
superconductivity but is broken in the presence of interlayer interactions.
In this work, we extend this approach to triplet pairs. We propose
interlayer triplet-pairing operators for the multi-layer extended Hubbard
model. We find that a set of exact condensate-pair eigenstates can be
constructed, which exhibit off-diagonal long-range order. In contrast to the 
$\eta$-pairing mechanism, this originates from restricted spectrum
generating algebra and is only available for bilayer and trilayer systems in
the presence of interlayer Hubbard interactions. Nevertheless, the system
also retains the original on-site $\eta$-pairing symmetry in the absence
of interlayer interactions. Consequently, both singlet and triplet pairs
coexist in the eigenstates of the multi-layer Hubbard model. We employ
quench dynamics to demonstrate the results through numerical simulations.
Our findings open avenues for the study of exact condensate-pair states in
strongly correlated systems.
\end{abstract}
\maketitle

\section{Introduction}

$\eta$-pairing state represents a specialized, exotic superconducting
state where pairs of on-site electrons form in simple Hubbard model on a
bipartite lattice \cite{Yang89}. The existence of $\eta$-pairing states in
the fermionic Hubbard model is fundamentally rooted in the $\eta$-pairing
symmetry of the Hubbard Hamiltonian \cite{YangZhang90, Pernici90}. One of
the most compelling findings is the potential for superconductivity, which
arises due to the off-diagonal long-range order (ODLRO) exhibited by these
states. This has led to the development of various non-equilibrium
protocols, such as external field-driven \cite{Rosch08, Kantian10, Kaneko19,
Kaneko20, Ejima20, Werner19, Li20, Kitamura16, Peronaci20, Cook20,
Tindall21, Tindall21_2, Diehl08, Kraus08, Bernier13, Buca19, Tindall19,
Tsuji21, Nakagawa21, Murakami21,zhang2022steady}, photodoping strategies 
\cite%
{iwai2003,Rosch08,sensarma2010,eckstein2011,ichikawa2011,lenarvcivc2013,stojchevska2014,mitrano2014,werner2019,Peronaci20,Li20}
and dissipation-based methods \cite%
{Diehl08,Kraus08,coulthard2017,werner2019,zhang2020b,zhang2021eta,yang2022dynamic}%
, aimed at selectively generating these superconducting-like states.
Recently, the $\eta$-pairing\ states in systems with edge states \cite%
{zhang2021topologically} and on a moire lattice \cite{wang2024flat} have
also been studied.

Unlike conventional BCS pairs, $\eta$-pairing provides a theoretical
example of unconventional superconductivity in strongly correlated systems.
On the other hand, multilayer Hubbard models are crucial theoretical
frameworks for studying high-temperature superconductivity in materials with
complex, multi-layered electronic structures. They extend the single-band
Hubbard model by introducing multiple planes or layers, typically with an
added interlayer hopping term and density-density interactions, which allow
for complex interactions and layer differentiation, especially in cuprates,
nickelates, and bilayer moire systems. Such interlayer coupling introduces
new degrees of freedom that can dramatically alter the phase diagram. For
instance, the two-dimensional bilayer Hubbard model has been extensively
investigated for their superconductivity, exciton condensation, magnetic
properties etc. \cite%
{bilayer1,bilayer2,bilayer3,bilayer4,bilayer5,bilayer6,bilayer7}.
Theoretically, interlayer density-density interactions break the $\eta$%
-pairing symmetry while preserving the SU(2) symmetry. Consequently, an $%
\eta$-pairing state ceases to be an eigenstate of the system, making an
alternative pairing mechanism specific to the multi-layered structure
desirable.

In this work, we investigate possible pairing eigenstates in the
multi-layered Hubbard model, which serve a function analogous to $\eta$%
-pairing states. We propose interlayer triplet-pairing operators rather than
on-site singlet-pairing operators for the multi-layer extended Hubbard
model. A set of exact condensate-pair eigenstates is constructed, exhibiting
off-diagonal long-range order (ODLRO). Unlike the $\eta$-pairing
mechanism, which originates from a spectrum-generating algebra (SGA) rooted
in symmetry, this arises from a restricted spectrum-generating algebra
(RSGA) \cite{moudgalya2020eta}. Notably, such eigenstates only exist in
bilayer and trilayer systems with interlayer Hubbard interactions.
Furthermore, the system retains the original on-site $\eta$-pairing
symmetry when interlayer interactions are absent. Consequently, singlet and
triplet pairs coexist in the eigenstates of the multi-layer Hubbard model.
We employ quench dynamics to demonstrate the results through numerical
simulations. Experimentally, new developments in quantum simulations of the
Hubbard model with ultracold atoms have given us a powerful tool to study
the low-temperature properties of strongly correlated systems \cite%
{bakr2009,parsons2015,cheuk2015,cheuk2016,parsons2016,esslinger2010}. This
may provide a way to measure the results obtained in this work in
experiment. Our findings open avenues for the study of exact condensate-pair
states in strongly correlated systems.

This paper is organized as follows. In Sec. ~\ref{Model and pseudospin
operators}, we introduce the multi-layer extended Hubbard model and the
relevant pseudospin operators. Sec. ~\ref{Triplet pairing eigenstates}
focuses on the construction of the triplet-pairing eigenstates. Sec. ~\ref%
{Stability of triplet-pair condensate} is dedicated to numerical simulations
of the quench dynamics. Finally, Sec. ~\ref{Summary} provides a summary of
the key findings and conclusions of this study.

\begin{figure}[t]
\centering
\includegraphics[width=0.52\textwidth]{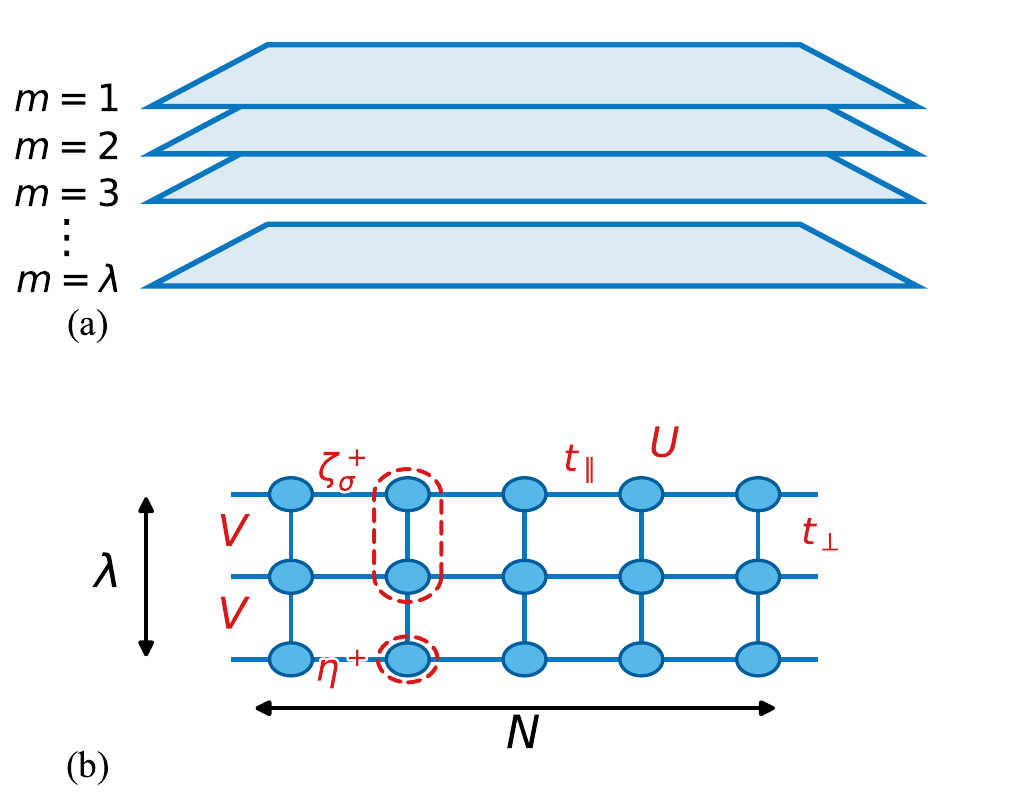}
\caption{(a) Schematic illustration of a multi-layered lattice system. The
index $m\in \left[ 1,\protect\lambda \right] $ labels the layers. All layers
are identical bipartite lattices. The interlayer hopping and interaction are
nearest-neighbor. (b) The simplest example of the system is an $N\times 
\protect\lambda $ square lattice, where the layers reduce to chains. The
parameters of the extended Hubbard model are indicated in the panel. The key
feature is that the nearest-neighbor interactions exist only between layers.
The triplet pair presented in this work is formed across two
nearest-neighbor sites on two layers (labeled by a red dashed line
surrounding the two sites), while the singlet pair is an on-site pair
(labeled by a red dashed line surrounding a single site). The main
conclusion of this work is that the multi-triplet pairing states can be
exact eigenstates for systems with nonzero $U$ and $V$, restricted to $%
\protect\lambda =2$ and $3$.}
\label{Fig1}
\end{figure}

\section{Model and pseudospin operators}

\label{Model and pseudospin operators}

The extended Hubbard model is one of the central theoretical models in
condensed matter physics for studying interacting fermions on lattices
beyond the simplest local-interaction approximation. It enables the
description of richer collective phenomena such as charge ordering,
unconventional superconductivity, and competing quantum phases. In this
work, we only consider an extension of the standard Hubbard model that
incorporates interlayer interactions.

We consider the $\lambda$-layer extended Hubbard model on a two-dimensional
square lattice with the Hamiltonian 
\begin{align}
H& =-t\sum_{\langle ij\rangle ,\sigma ,m}c_{i,m,\sigma }^{\dagger
}c_{j,m,\sigma }-t_{\perp }\sum_{i,\sigma ,m=1}c_{i,m,\sigma }^{\dagger
}c_{i,m+1,\sigma }  \notag \\
&\quad +\mathrm{H.c.}+U\sum_{i,m}n_{i,m,\uparrow }n_{i,m,\downarrow
}+V\sum_{i,m=1}^{\lambda -1}n_{i,m}n_{i,m+1},  \label{H}
\end{align}%
where $c_{i,m,\sigma }^{\dagger }$ ($c_{i,m,\sigma }$) creates (annihilates) an
electron with spin $\sigma $ ($\sigma =\uparrow $,$\downarrow $) in $m$-th ($%
m=1,2,...\lambda $) layer at site $i$. The particle number operators are
defined by $n_{i,m,\sigma }=c_{i,m,\sigma }^{\dagger }c_{i,m,\sigma }$ and $%
n_{i,m}=\sum_{\sigma =\uparrow ,\downarrow }n_{i,m,\sigma }$, respectively.\
The intralayer (interlayer) nearest-neighbor (NN) hopping strength is $t$ ($%
t_{\perp }$). The on-site interaction strength is $U$, the interlayer NN
interaction strength is $V$. Fig. \ref{Fig1} shows a schematic illustration
of the model. Although we focus on square lattices for each layer, our
conclusion holds as long as each layer has a bipartite lattice geometry.

We start with the symmetry of the Hamiltonian $H$ given in Eq. (\ref{H}). We
introduce three sets of operators that obey the SU(2) Lie algebra and play
an important role in this work.

(i) Spin operators. The spin operators are defined as%
\begin{eqnarray}
s^{+} &=&\left( s^{-}\right) ^{\dagger }=\sum_{j,m}c_{j,m,\uparrow
}^{\dagger }c_{j,m,\downarrow }, \\
s^{z} &=&\frac{1}{2}\sum_{j,m}\left( n_{j,m,\uparrow}-n_{j,m,\downarrow
}\right) ,  \label{pseudo spin}
\end{eqnarray}%
which obey the SU(2) Lie algebra, i.e., $[s^{+},s^{-}]=2s^{z}$, and $%
[s^{z},s^{\pm }]=\pm s^{\pm }$. For arbitrary $U$ and $V$, we have%
\begin{equation}
\left[ H,s^{\pm }\right] =\left[ H,s^{z}\right] =0.  \label{su2 symm}
\end{equation}%
The full Hilbert space decomposes into sectors labeled by the eigenvalues of 
$s^{z}$. Using the SGA, the operators $s^{\pm }$ generate non-zero
eigenstates $\left( s^{\pm }\right) ^{n}\left\vert \phi _{0}\right\rangle $
in different sectors from an arbitrary eigenstate $\left\vert \phi
_{0}\right\rangle $. In general, $s^{\pm }\left\vert \phi _{0}\right\rangle
=0$\ if $\left\vert \phi _{0}\right\rangle $\ is a singlet state. In our
work, however, $\left\vert \phi _{0}\right\rangle $\ is a triplet pairing
state, so that $s^{\pm }\left\vert \phi _{0}\right\rangle \neq 0$.

(ii) Singlet $\eta$-pairing operators. The singlet $\eta$-pairing
operators are defined as%
\begin{eqnarray}
\eta ^{+} &=&\left( \eta ^{-}\right) ^{\dagger }=\sum_{j,m}\left( -1\right)
^{j+m}c_{j,m,\uparrow }^{\dagger }c_{j,m,\downarrow }^{\dagger }, \\
\eta ^{z} &=&\frac{1}{2}\sum_{j,m}\left( n_{j,m,\uparrow }+n_{j,m,\downarrow
}-1\right) ,
\end{eqnarray}%
which obey the SU(2) Lie algebra, i.e., $[\eta ^{+},\eta ^{-}]=2\eta ^{z}$,
and $[\eta ^{z},\eta ^{\pm }]=\pm \eta ^{\pm }$. For $V=0$, it has been
shown that%
\begin{equation}
\left[ H,\eta ^{\pm }\right] =\pm U\eta ^{\pm },  \label{eta symm}
\end{equation}%
which is known as $\eta$-pairing symmetry.\ This symmetry allows us to
generate nonzero eigenstates of the form $\left( \eta ^{\pm }\right)
^{n}\left\vert \phi _{0}\right\rangle $\ from an arbitrary eigenstate $%
\left\vert \phi _{0}\right\rangle $\ via the SGA. These are referred to as $%
\eta $-pairing states. We note that when $\left\vert \phi _{0}\right\rangle $%
\ is the vacuum state, it\ is a singlet state, so that $s^{\pm }\left\vert
\phi _{0}\right\rangle =0$.

(iii) Triplet $\zeta$-pairing operators. They are the key operators in this
work, defined as%
\begin{eqnarray}
\zeta _{\sigma }^{+} &=&\left( \zeta _{\sigma }^{-}\right) ^{\dagger }\notag \\
&=&\sum_{j=1}^{N}\sum_{m=1}^{\lambda -1}(-1)^{j+m}c_{j,m,\sigma }^{\dagger
}c_{j,m+1,\sigma }^{\dagger } \\
\zeta _{\sigma }^{z} &=&\frac{1}{2}\sum_{j=1}^{N}[\sum_{m=1}^{\lambda
-1}\left( n_{j,m,\sigma }+n_{j,m+1,\sigma }-1\right)  \notag \\
&&+\sum_{m=1}^{\lambda -2}\left( c_{j,m,\sigma }^{\dagger }c_{j,m+2,\sigma
}+c_{j,m+2,\sigma }^{\dagger }c_{j,m,\sigma }\right) ],
\end{eqnarray}%
with $\sigma =\uparrow $,$\downarrow $, which also obey the SU(2)\ Lie
algebra, i.e., $[\zeta _{\sigma }^{+},$ $\zeta _{\sigma ^{\prime
}}^{-}]=2\zeta _{\sigma }^{z}\delta _{\sigma \sigma ^{\prime }}$, and $%
[\zeta _{\sigma }^{z},$ $\zeta _{\sigma ^{\prime }}^{\pm }]=\pm \zeta
_{\sigma }^{\pm }\delta _{\sigma \sigma ^{\prime }}$. Compared to the
operators $s$\ and $\eta$, the $\zeta$ operators\ have subtle relations
with the Hamiltonian $H$. For the trivial cases with $V=U=0$, we have%
\begin{equation}
\left[ H,\zeta _{\sigma }^{\pm }\right] =0,
\end{equation}%
which also allows us to generate nonzero eigenstates of the form $\left(
\zeta _{\sigma }^{\pm }\right) ^{n}\left\vert \phi _{0}\right\rangle $\ from
an arbitrary eigenstate $\left\vert \phi _{0}\right\rangle $\ by the SGA. We
refer to such states as $\zeta$-pairing states. Unlike the $\eta$-pairing
state, they seem trivial since the system is a noninteracting system and
thus the NN pair has no binding energy. Nevertheless, we will show that such
operators $\zeta _{\sigma }^{\pm }$\ can also be utilized to generate new
eigenstates when $V$ and $U$\ are nonzero if the Hamiltonian satisfies
certain conditions. This is the main point of this work.

Moreover, the relations among the three sets of operators are obtained as%
\begin{equation}
\left[ s^{\alpha },\eta ^{\beta }\right] =0,\left[ \zeta _{\sigma }^{\alpha
},\eta ^{\beta }\right] \neq 0,\left[ \zeta _{\sigma }^{\alpha },s^{\beta }%
\right] \neq 0,
\end{equation}%
with $\alpha ,\beta =\pm ,z$, which will be used to investigate the
coexistence of two types of pairs.

\section{Triplet pairing eigenstates}

\label{Triplet pairing eigenstates}

In the preceding section, we demonstrated the existence of triplet-pairing
eigenstates based on the SGA for the noninteracting system. A natural
question is whether the system $H$ supports $\zeta$-pairing eigenstates in
the presence of nonzero $V$ and $U$. Recent results show that a set of
eigenstates can be constructed using fermionic-pair or hard-core-bosonic
operators without requiring any symmetry \cite%
{ma2022steady,zhang2025coalescing,HDK_PRB2025,LSJY_PRB1,LSJY_PRB2,
LSJY_PRB3,LFX_arXiv}. They motivate us to study the problem using the RSGA
framework. We answer this question using the following commutation relations 
\begin{eqnarray}
\left[ H,\zeta _{\sigma }^{+}\right]  &=&V\zeta _{\sigma
}^{+}-V\tsum\limits_{j=1}^{N}\tsum\limits_{m=1}^{\lambda -2}\left( -1\right)
^{j+m}c_{j,m+1,\sigma }^{\dagger }  \notag \\
&&\times (c_{j,m,\sigma }^{\dagger }n_{j,m+2,\sigma }+c_{j,m+2,\sigma
}^{\dagger }n_{j,m,\sigma })  \notag \\
&&+U\tsum\limits_{j=1}^{N}\tsum\limits_{m=1}^{\lambda -1}\left( -1\right)
^{j+m}c_{j,m,\sigma }^{\dagger }c_{j,m+1,\sigma }^{\dagger }  \notag \\
&&\times (n_{j,m,\sigma ^{\prime }}+n_{j,m+1,\sigma ^{\prime }}),
\label{comm}
\end{eqnarray}%
with $\sigma \neq \sigma ^{\prime }$, and the double commutator 
\begin{eqnarray}
&&\left[ \left[ H,\zeta _{\sigma }^{+}\right] ,\zeta _{\sigma }^{+}\right]  
\notag \\
&=&2V\tsum\limits_{j=1}^{N}\tsum\limits_{m=1}^{\lambda -3}c_{j,m,\sigma
}^{\dagger }c_{j,m+1,\sigma }^{\dagger }c_{j,m+2,\sigma }^{\dagger }c_{j,m+3,\sigma
}^{\dagger }.
\end{eqnarray}%
Here we focus on the operator $\zeta _{\sigma }^{+}$. The corresponding
relation for $\zeta _{\sigma }^{-}$ can be obtained by taking the Hermitian
conjugate of the above relation. We find that the commutator depends not only
on $V$ and $U$, but also on $\lambda$. This indicates that the original
Hamiltonian in its general form does not satisfy the RSGA conditions. In what
follows, we restrict the Hamiltonian parameters to meet the RSGA conditions.

Acting with the commutator in Eq. (\ref{comm}) on the vacuum state $\left\vert
0\right\rangle $, which is an eigenstate of $H$, we have%
\begin{equation}
\left[ H,\zeta _{\sigma }^{+}\right] \left\vert 0\right\rangle =V\zeta
_{\sigma }^{+}\left\vert 0\right\rangle .
\end{equation}%
Furthermore, we have%
\begin{equation}
\left[ \left[ H,\zeta _{\sigma }^{+}\right] ,\zeta _{\sigma }^{+}\right] =0,
\end{equation}%
only for $\lambda =2$ and $3$. These conditions guarantee that RSGA applies,
yielding a set of degenerate eigenstates%
\begin{equation}
\left\vert \phi _{n}\right\rangle =\left( \zeta _{\sigma }^{+}\right)
^{n}\left\vert 0\right\rangle .
\end{equation}%
This behavior is distinct from all previous scenarios. The mechanism
underlying the restriction on $\lambda$ for the existence of the state $\left\vert
\phi _{n}\right\rangle $ is as follows. For the $2$- or $3$-layer system,
there is at most one NN pair associated with the sites with the same $j$ (the
vertical dimer or trimer) in the eigenstates. When $V$ is switched on, it does
not contribute to an inter-pair interaction. By contrast, for $\lambda >3$,
the nonzero inter-pair interaction destroys these eigenstates.

\begin{table*}[t]
	\caption{The explicit form of eigenstates for the system with parameters $U$%
		, $V$, and the number of the layers $\protect\lambda $.}%
	\begin{ruledtabular}
		\begin{tabular}{llll}
			$U$ & zero & nonzero & nonzero \\ 
			$V$ & nonzero & zero & nonzero \\ 
			$\lambda =2,3$ & $\left( s^{\pm }\right) ^{n}\left( \zeta _{\sigma
			}^{+}\right) ^{k}\left\vert 0\right\rangle $ & $\left( s^{\pm }\right)
			^{n}\left( \eta ^{+}\right) ^{m}\left( \zeta _{\sigma }^{+}\right)
			^{k}\left\vert 0\right\rangle $ & $\left( s^{\pm }\right) ^{n}\left( \zeta
			_{\sigma }^{+}\right) ^{k}\left\vert 0\right\rangle $ \\ 
			$\lambda >3$ & none & $\left( s^{\pm }\right) ^{n}\left( \eta ^{+}\right)
			^{m}\left( \zeta _{\sigma }^{+}\right) ^{k}\left\vert 0\right\rangle $ & none%
		\end{tabular}%
	\end{ruledtabular}
\end{table*}

Considering the Hamiltonian on $N\times N$ square lattice with $\lambda =2$
and $3$, the corresponding eigenstate can be explicitly written as%
\begin{equation}
\left\vert \psi _{n}\right\rangle =\frac{1}{n!\sqrt{C_{N^{2}}^{n}(\lambda
-1)^{n}}}\left( \zeta _{\sigma }^{+}\right) ^{n}\left\vert 0\right\rangle ,
\end{equation}%
where the pair operator is in the form

\begin{equation}
\zeta _{\sigma }^{+}=\sum_{\mathbf{r}=(1,1)}^{(N,N)}\sum_{m=1}^{\lambda
-1}e^{i\mathbf{\pi }\cdot \mathbf{r}}\left( -1\right) ^{m}c_{\mathbf{r}%
,m,\sigma }^{\dagger }c_{\mathbf{r},m+1,\sigma }^{\dagger },
\end{equation}%
with $\mathbf{\pi }=(\pi ,\pi )$. Straightforward derivation shows the
correlation function as%
\begin{eqnarray}
&&\left\langle \psi _{n}\right\vert c_{\mathbf{r},1,\sigma }^{\dagger }c_{\mathbf{r},2,\sigma }^{\dagger }
c_{\mathbf{r}^{\prime },1,\sigma }c_{\mathbf{r}^{\prime },2,\sigma }\left\vert \psi _{n}\right\rangle \notag \\
&=&-\frac{e^{i\mathbf{\pi }\cdot \left( \mathbf{r}-\mathbf{r}^{\prime }\right) }\left( N^{2}-n\right) n}{(\lambda -1)N^{4}}.
\label{<cccc>}
\end{eqnarray}%
This shows that the state $\left\vert \psi _{n}\right\rangle $ possesses
ODLRO, as shown in Ref. \cite{yang1962concept}, because the correlation
function does not decay with increasing $\left\vert \mathbf{r}-\mathbf{r}^{\prime
}\right\vert $. The result stems from the fact that the amplitudes of $c_{%
\mathbf{r},m,\sigma }^{\dagger }c_{\mathbf{r},m+1,\sigma }^{\dagger }$\ in
the operator $\zeta _{\sigma }^{+}$\ is position independent up to a $\pm $\
sign. Consequently, the correlation function is also position independent up
to a $\pm $\ sign. A similar conclusion can be obtained for the
correlations $\left\langle \psi _{n}\right\vert c_{\mathbf{r},2,\sigma }^{\dagger }c_{%
\mathbf{r},3,\sigma }^{\dagger }c_{\mathbf{r}^{\prime },2,\sigma }c_{\mathbf{r}^{\prime
},3,\sigma }\left\vert \psi _{n}\right\rangle $ and $%
\left\langle \psi _{n}\right\vert c_{\mathbf{r},2,\sigma }^{\dagger }c_{\mathbf{r}%
,3,\sigma }^{\dagger }c_{\mathbf{r}^{\prime },1,\sigma }c_{\mathbf{r}^{\prime },2,\sigma }\left\vert
\psi _{n}\right\rangle $.

On the other hand, based on these eigenstates, many other eigenstates can be
obtained by using the relations in Eqs. (\ref{su2 symm}) and (\ref{eta symm}%
), including the ones with the coexistence of two types of pairs. We
summarize the obtained eigenstates in Table I.

Throughout the previous investigation, we have only considered the triplet
pair states with aligned spin. This can simplify our description. In fact,
the triplet pair operator can take the form 
\begin{eqnarray}
\zeta _{0}^{+} &=&\left( \zeta _{0}^{-}\right) ^{\dagger }\notag \\
&=&\sum_{j=1}^{N}\sum_{m=1}^{\lambda -1}(-1)^{j+m}  \notag \\
&&\times \frac{1}{\sqrt{2}}(c_{j,m,\uparrow }^{\dagger }c_{j,m+1,\downarrow
}^{\dagger }+c_{j,m,\downarrow }^{\dagger }c_{j,m+1,\uparrow }^{\dagger }),
\end{eqnarray}%
which can be shown to play the same role as $\zeta _{\sigma }^{\pm }$. This
is due to the fact that the Hamiltonian has SU(2) symmetry, given by Eq. (%
\ref{su2 symm}).

\section{Stability of triplet-pair condensate}

\label{Stability of triplet-pair condensate}

Based on the analysis in the previous section, the triplet-pairing states are
eigenstates of the system for any $\lambda >1$ in the absence of NN interaction.
However, they are no longer eigenstates of the system for $\lambda >3$ in the
presence of NN interaction. In this section, we demonstrate the stability of
the triplet-pair condensate by quench dynamics. The initial states are
triplet-pair condensate states of the $\lambda$-layer Hubbard model with zero
$V$ and nonzero $U$. The post-quench Hamiltonian has nonzero $V$ and $U$.
Specifically, we consider the time evolution of the triplet-pair condensate as
the initial state under the Hamiltonian with different values of $V$ and $\lambda$.
The geometry of the system is simplified by a $\lambda$-leg ladder, or $N\times \lambda $\ square
lattice. The evolved states are%
\begin{equation}
\left\vert \phi \left( t\right) \right\rangle =e^{-iHt}\left\vert \psi
_{n}\right\rangle ,
\end{equation}%
which can be obtained by exact diagonalization for finite systems. It is
expected that the evolved state remains unchanged when $\lambda =2$ and $3$
but deviates from the initial state otherwise. To quantify the dynamic
response, we use the fidelity%
\begin{equation}
F(t)=\left\vert \left\langle \psi _{n}\right\vert e^{-iHt}\left\vert \psi
_{n}\right\rangle \right\vert ^{2},  \label{F}
\end{equation}%
which characterizes how the state departs from its initial state as a function
of the NN interaction strength $V$ for a given $\lambda$.

\begin{figure}[!t]
\centering
\includegraphics[width=0.9\columnwidth]{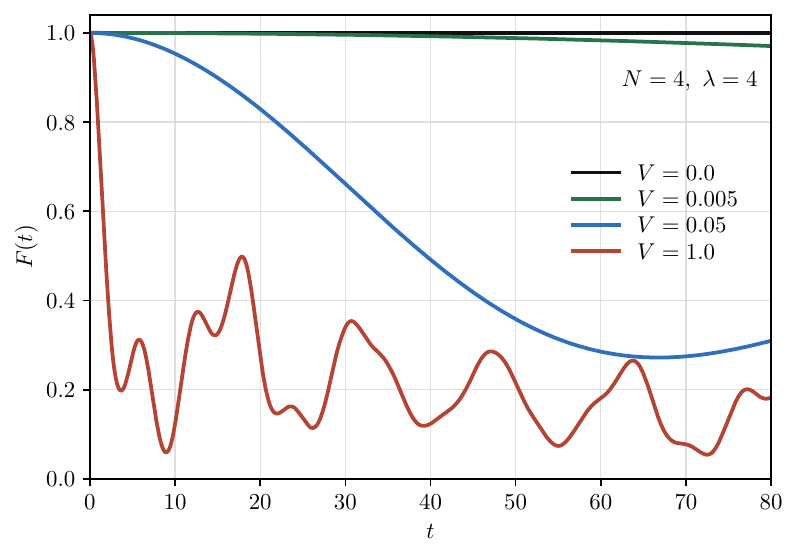}
\caption{Plots of the fidelity defined in Eq. (\protect\ref{F}) for the
initial triplet-pair state $\left( \protect\zeta _{\uparrow }^{+}\right)
^{4}\left\vert 0\right\rangle $ under the Hamiltonian on the $4\times 4$
lattice. The four curves correspond to several typical values of $V$. The other
parameters for the Hamiltonian are $t=1$, $U=1$, and $N=4$. The results show
that $F(t)$ remains unity when $V=0$, while finite $V$ drives the state away
from the initial triplet-pair state for $\lambda =4$. This indicates that,
unlike for the systems on $4\times 2$ and $4\times 3$ lattices, the state is
no longer an eigenstate of the system on $4\times 4$ when $V$ is nonzero, in
accordance with our predictions.}
\label{fig2}
\end{figure}

As shown above, we have $F(t)=1$ for all times when $\lambda =2$ and $3$. We
plot $F(t)$ in Fig. \ref{fig2} as a function of $t$ for selected system sizes,
particle numbers, and $V$. The results accord with our predictions. The
$F(t)$ remains unchanged for the cases with zero $V$, or nonzero $V$ but
$\lambda =2$ and $3$. In contrast, $F(t)$ decays with time $t$ for nonzero
values of $V$ and $\lambda >3$ as expected.

\begin{figure*}[!t]
	\centering
	\includegraphics[width=0.90\textwidth]{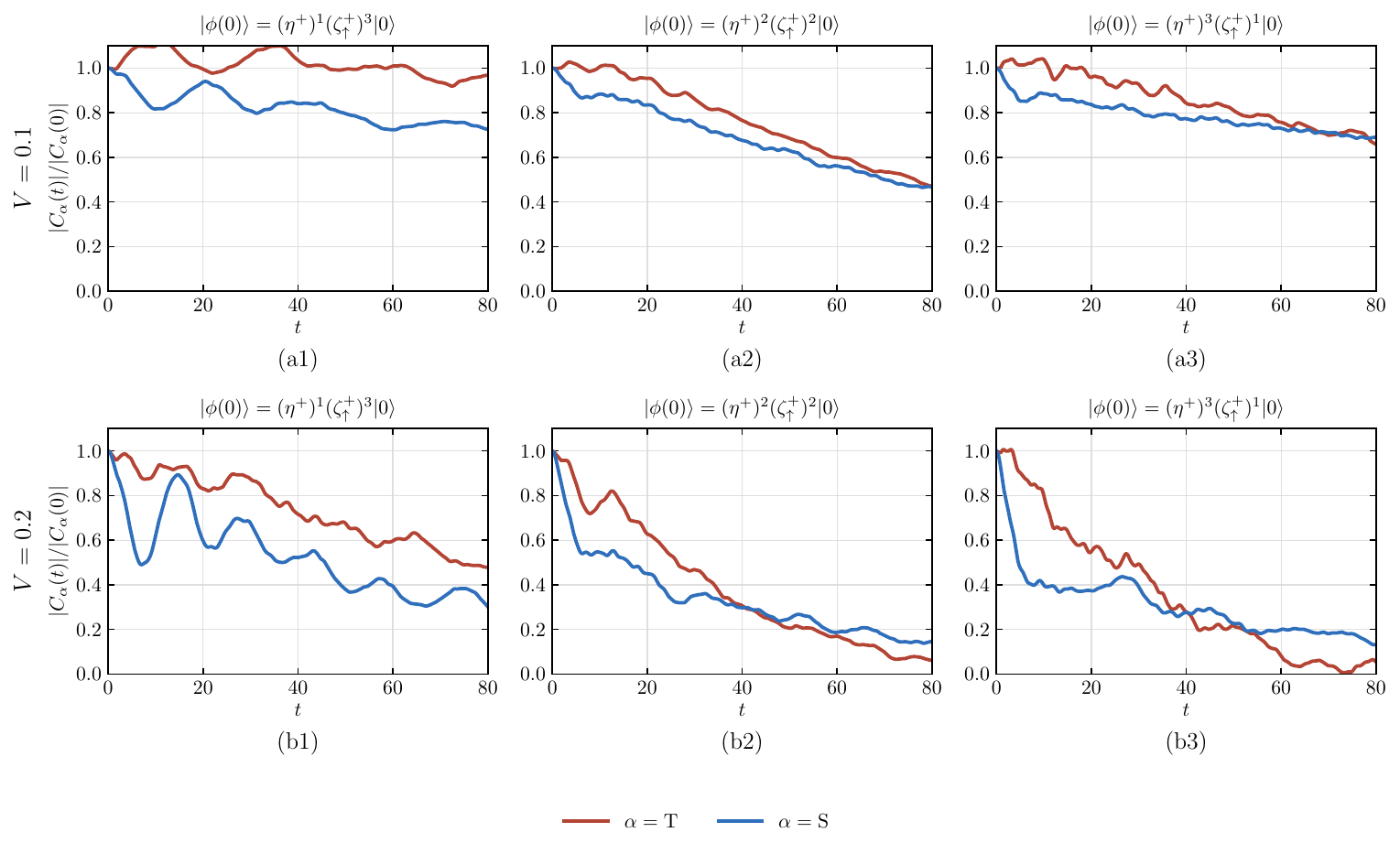}
	\caption{Plots of the two correlation functions $C_{\text{S}}(r,t)$\ and $C_{%
			\text{T}}(r,t)$ defined in Eqs.~(\protect\ref{CS}) and (\protect\ref{CTCS}) for three initial pair
		states $\protect\eta ^{+}\left( \protect\zeta _{\uparrow }^{+}\right)
		^{3}\left\vert 0\right\rangle $, $\left( \protect\eta ^{+}\right) ^{2}\left( 
		\protect\zeta _{\uparrow }^{+}\right) ^{2}\left\vert 0\right\rangle $, and $%
		\left( \protect\eta ^{+}\right) ^{3}\protect\zeta _{\uparrow }^{+}\left\vert
		0\right\rangle $,\ on a $4\times 3$ lattice. The two points used for the
		correlation functions are diagonally opposite points, ($1,1$) and ($4,3$),
		on the $4\times 3$ square lattice. The values of $V$ are indicated in the
		panels. The other parameters are $t=1$ and $U=1$. We have shown that both $C_{%
			\text{S}}(r,t)$\ and $C_{\text{T}}(r,t)$ remain unchanged when $V=0$,
		regardless of the value of $U$. In contrast, all the panels show that both $%
		C_{\text{T}}(r,t)$\ and\ $C_{\text{S}}(r,t)$ change when $V$ is finite. Two
		factors influence the change of the two correlation functions: (i) the
		numbers of the $\protect\eta $\ pairs; (ii) the values of $V$. A small
		number of $\protect\eta $\ pairs\ and a small value of $V$ result in
		relatively slower decay of the correlation functions. It indicates that the
		presence of singlet pairs leads to the instability of the triplet pairs.}
	\label{fig3}
\end{figure*}

On the other hand, we know that nonzero $V$ destroys the $\eta$-pairing
states. We consider an initial state of the form $\left\vert \phi \left(
0\right) \right\rangle =\left( \eta ^{+}\right) ^{n}\left( \zeta _{\sigma
}^{+}\right) ^{m}\left\vert 0\right\rangle $, in which two types of pairing
eigenstates coexist when $V=0$. We employ two types of correlation functions%
\begin{eqnarray}
C_{\text{S}}(r,t) &=&\left\langle \phi \left( t\right) \right\vert
c_{j,1,\uparrow }^{\dagger }c_{j,1,\downarrow }^{\dagger }c_{j+r,1,\uparrow
}c_{j+r,1,\downarrow }\left\vert \phi \left( t\right) \right\rangle ,
\label{CS}
\\
C_{\text{T}}(r,t) &=&\left\langle \phi \left( t\right) \right\vert
c_{j,1,\uparrow }^{\dagger }c_{j,2,\uparrow }^{\dagger }c_{j+r,1,\uparrow
}c_{j+r,2,\uparrow }\left\vert \phi \left( t\right) \right\rangle ,
\label{CTCS}
\end{eqnarray}%
to examine the stability of the two types of pair-condensate states. This
state contains $n$ $\eta$-pairs and $m$ $\zeta$-pairs, and is an eigenstate of
$H$ with nonzero $U$, zero $V$, and $\lambda <4$. We compute the two correlation
functions $C_{\text{S}}(r,t)$ and $C_{\text{T}}(r,t)$ numerically for nonzero
$V$. As shown above, a nonzero interlayer interaction $V$ breaks the $\eta$-pairing symmetry, whereas the pure $\zeta$-pairing states remain exact eigenstates for $\lambda=2$ and $3$. In a state where the two types of pairs coexist, however, the singlet and triplet pairing correlations may become coupled, since the corresponding pair-creation operators $\eta^{+}$ and $\zeta_{\uparrow}^{+}$ do not commute.

Both $C_{\text{S}}(r,t)$ and $C_{\text{T}}(r,t)$ remain unchanged in the absence
of $V$. In contrast, the results show that both $C_{\text{T}}(r,t)$ and $C_{%
\text{S}}(r,t)$ decay with time $t$ for nonzero values of $V$. Nonzero $V$ does
not affect triplet pairs directly, but the latter may be destroyed indirectly
by the breaking of singlet pairs, since the operators $\eta ^{+}$ and $\zeta
_{\uparrow }^{+}$ do not commute. To examine the effect of singlet pairs on
the triplet pairs, we compute the cases with $(m,n)=(1,4)$, $(2,2)$, and
$(3,1)$. The results show that the
existence of singlet pairs\ indeed destroys triplet pairs.\ The two types of
pairs cannot coexist in the presence of $V$.\ Such a system supports
superconducting states based on either singlet pairs\ or triplet pairs.

\section{Summary}\label{Summary} 

In summary, we have extended the singlet $\eta$-pairing
state to the triplet $\zeta$-pairing state in the extended Hubbard model
on the multi-layer square lattice. We have shown that a set of exact
condensate-pair eigenstates exhibiting off-diagonal long-range order can be
constructed based on the RSGA. In contrast to the $\eta$-pairing
mechanism, such triplet-pair states only exist in the bilayer and trilayer
systems in the presence of interlayer Hubbard interactions. This indicates
that such triplet-pair states may replace singlet pairs as a possible
mechanism of superconductivity in a quasi-2D square-lattice Hubbard system.
Numerical simulations have been performed to demonstrate the above
conclusions. The proposed set of triplet-pair eigenstates provides a
framework for studying exact condensate-pair states in strongly correlated
systems.

\section*{ACKNOWLEDGMENTS}

We acknowledge the support of NSFC (Grants No. 12374461).

\makeatletter
\let\pre@bibdata\@empty
\makeatother
\providecommand{\enquote}[1]{#1} 
\bibliography{reference}

\end{document}